\documentclass[structabstract]{aa}
\usepackage{txfonts}
\usepackage{natbib}
\usepackage{graphicx}
\usepackage{longtable,lscape}
\bibpunct{(}{)}{;}{a}{}{,} 

\begin{document}

\title{Is the far border of the Local Void expanding?
  \thanks{Based on data taken at Nan\c{c}ay radiotelescope operated by 
  Observatoire de Paris, CNRS and Universit\'{e} d'Orl\'{e}ans, 
  Infrared Survey Facility (IRSF) which is operated by Nagoya university
  under the cooperation of South African Astronomical Observatory, 
  Kyoto University, and National Astronomical Observatory of Japan.
  }
\fnmsep
  \thanks{
  This publication makes use of data products from the Two Micron All Sky
  Survey, which is a joint project of the University of Massachusetts
  and the Infrared Processing and Analysis Center/California Institute
  of Technology, funded by the National Aeronautics and Space
  Administration and the National Science Foundation.
}}

\author{
 Ikuru Iwata\inst{1}
  \thanks{\emph{Present address: Subaru Telescope, National Astronomical
 Observatory of Japan, 650 North A'ohoku Place, Hilo, HI 96720 U.S.A.}}
  \and
Pierre Chamaraux\inst{2}
   }

\offprints{I. Iwata, \email{ ikuru.iwata@nao.ac.jp}}

\institute{
  Okayama Astrophysical Observatory, 
  National Astronomical Observatory of Japan, 
  Honjo, Kamogata, Asakuchi, Okayama 719-0232, Japan
  \and 
  Observatoire de Paris - Meudon, GEPI, 92195 Meudon Cedex, France
}

\date{Received 10 February 2011 / Accepted 14 April 2011}

\abstract {According to models of evolution in the hierarchical
structure formation scenarios, voids of galaxies are expected to 
expand. The Local Void (LV) is the closest large void, and it provides a 
unique opportunity to test observationally such an expansion. It has 
been found that the Local Group, which is on the border of the LV, is 
running away from the void center at $\sim$260 km s$^{-1}$.}{In this 
study we investigate the motion of the galaxies at the far-side border 
of the LV to examine the presence of a possible expansion.}{We selected 
late-type, edge-on spiral galaxies with radial velocities between 3000 km 
s$^{-1}$ and 5000 km s$^{-1}$, and carried out HI 21 cm line and
$H$-band imaging observations. The near-infrared Tully-Fisher relation
was calibrated with a large sample of galaxies and carefully
corrected for Malmquist bias. It was used to compute the
distances and the peculiar velocities of the LV sample galaxies. 
Among the 36 sample LV galaxies with
good quality HI line width measurements, only 15 galaxies were 
selected for measuring their distances and peculiar velocities, in order
to avoid the effect of Malmquist bias.}
{The average peculiar velocity of these 15 galaxies is found to be 
$-419^{+208}_{-251}$ km s$^{-1}$,
which is not significantly different from zero.}{Due to the
intrinsically large scatter of Tully-Fisher relation, we cannot conclude
whether there is a systematic motion against the center of the LV for
the galaxies at the far-side boundary of the void. However, our result
is consistent with the hypothesis that those galaxies at the far-side
boundary have an average velocity of $\sim$260 km s$^{-1}$ equivalent to
what is found at the position of the Local Group.}


\maketitle 

\section{Introduction}

Deep extended galaxy surveys have shown that the large-scale
distribution of galaxies consists in matter concentrations, such as
clusters, filaments, and walls, and also in vast regions devoid of
galaxies, i.e. the voids. These voids occupy the largest
volumes in the Universe, according to 
\citet{ceccarelli2006}, and 
the radii of the voids those authors find in the 2dF galaxy
redshift survey range from 5 to 25 $h^{-1}$ Mpc ($h = H_0/100$). 

Voids are expected to expand, since galaxies undergo a gravitational
pull at their borders from the objects located outside them.  
\citet{sheth2004} 
have developed a model of the evolution of voids, which
indeed leads to an expansion of the surviving voids at the present
time. On the other hand, \citet{ceccarelli2006} 
have modeled the velocity field around the voids found in their
study, and show that the expansion velocity is maximum at the edge of
the voids and is proportional to the void radius, for instance reaching 
210 km s$^{-1}$ for a void with a radius of $12.5 h^{-1}$ Mpc. 

It is also possible to directly measure the expansion velocities at the
edge of a peculiar void, namely the Local Void, taking advantage of its
being very close to us. The Local Void (hereafter LV) was discovered 
by \citet{tully1987} 
from their survey of galaxies
with redshift lower than 3000 km s$^{-1}$. Its structure has been
investigated by \citet{nakanishi1997} 
from a visual search of IRAS
galaxies behind the Milky Way, since the major part of this void is at
galactic latitude $|b| < 15\degr$. They localize its center at
$\ell=60\degr$, $b=-15\degr$, $cz = 2500$ km s$^{-1}$, and they find
that it extends to $cz=5000$ km s$^{-1}$. On the other hand, the Local
Group and neighboring galaxies are located at the boundary of the LV, as
shown by \citet{tully2008}.

By accurate measurements of distances of 200 galaxies within 10 Mpc
carried out with the Hubble Space Telescope, 
\citet{tully2008} 
find that the Local Group and its neighboring galaxies are
running away from the center of the LV with a velocity of 259 
km s$^{-1}$. This proves the expansion of the LV at our location and
also solves the problem of the so-called ``Local Velocity Anomaly''
appearing in the motion of the LG relative to the CMB 
\citep{faber1988, burstein2000}.

In the present study, we intend to determine the peculiar velocities of
galaxies located at the edge of the LV opposite to us ($cz \sim$
3000--5000 km s$^{-1}$) in order to check whether the LV also
undergoes an observable expansion in that region. The peculiar
velocities are computed from the distances of the galaxies measured by
means of the near-infrared Tully-Fisher relation (hereafter IRTFR) using 
near-infrared and HI 21-cm observations. 

The organization of the paper is the following. Section 2 presents the
sample of the LV galaxies observed. And then the IR and HI measurements
are described and the data of interest are given. 
In Sect. 3, the IRTFR in $H$-band is determined from a calibration sample
and corrected for Malmquist bias.
In Sect. 4 we compute the distances of the LV galaxies from
the IRTFR and derive their peculiar velocities after correction of the 
observed radial velocities from infall into some nearby mass
concentrations. Concluding remarks are given in Sect. 5.

\section{Observations}

\subsection{The sample selection}

We selected uniquely spiral galaxies for the HI and IR observations,
since the TF relation is only valid for them. These objects were chosen
as located at the edge of the LV opposite to us and slightly beyond, at
galactic coordinates: $30\degr < l < 70\degr$, $|b|<20\degr$ (i.e.,
around the North Supergalactic Pole), and with recession velocities 
$cz < 5000$ km s$^{-1}$ \citep[see the maps by][]{nakanishi1997}.
In addition to the galaxies in the literature
(most of them are listed in the UGC catalog \citep{nilson1973}),
we executed the redshift measurement observations for 
some galaxies that were discovered in a systematic
optical search by \citet{roman2000}. 
The observations were done in July and October 2000 using the New
Cassegrain Spectrograph attached to the 188 cm reflector of the Okayama 
Astrophysical Observatory, National Astronomical Observatory of Japan.
In Table \ref{tab:optsurvey} we list these new radial velocities. Among
these galaxies with new radial velocity measurements, those at 
$cz < 5000$ km s$^{-1}$ were added to the sample, except UGC~11417,
which does not satisfy the limit of the axial ratio (less than 0.71; see
below).

Use of the TF relation needs the determination of the maximal rotational
velocity $V_m$ of each galaxy, and $V_m$ is obtained from the width $W$
of the 21-cm HI profile by 
$V_m \sim W / (2 \mathrm{sin} i)$, 
where $i$ is the inclination of the galaxy. To obtain accurate
$V_m$, we only kept galaxies with $i>45\deg$, i.e., galaxies with
axial ratios in the three-band coadded images in the Two Micron All Sky
Survey \citep[2MASS;][]{Skrutskie2006}, named ``sup-ba'' in the
extended source catalog (XSC) less than 0.71 (assuming an intrinsic
axis ratio of 0.2 for a spiral galaxy viewed edge-on). 

Moreover, we need good S/N in the HI profiles to obtain good HI
widths, hence accurate distances by the TF relation. Since the HI fluxes
of the spiral galaxies are proportional to the square of their apparent
diameters, such an accuracy is obtained by selecting only galaxies with
sufficiently large apparent diameters. Taking into account the
sensitivity of the Nan\c{c}ay radiotelescope, we kept mainly galaxies having
an extinction-corrected major axis larger than one arcmin. 

\begin{figure*}
\resizebox{\textwidth}{!}{\includegraphics{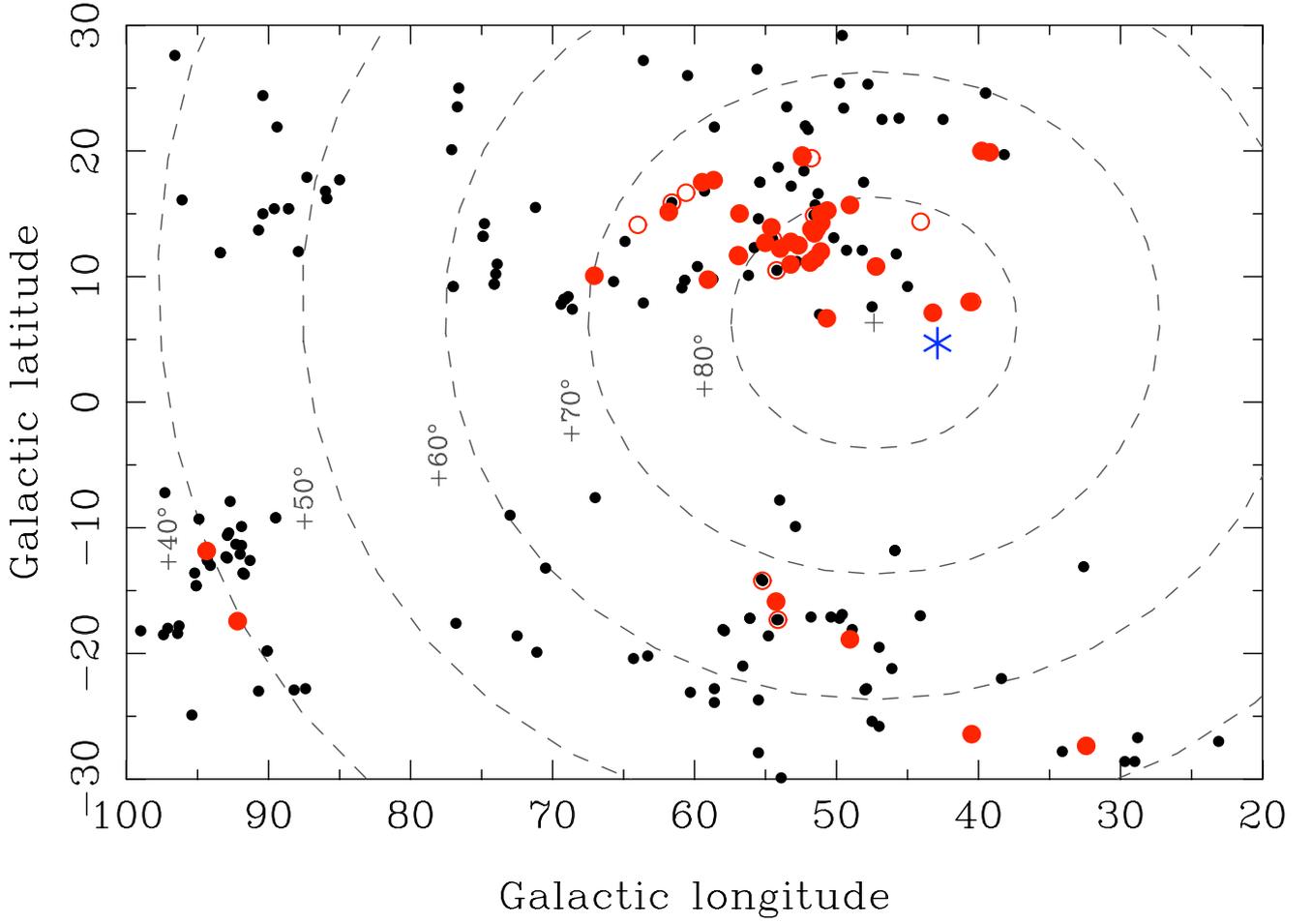}} 
\caption{Spatial distribution of the Local Void sample galaxies in 
 galactic coordinates. Small
 filled circles represent galaxies in the IRAS PSCz catalog 
\citep{saunders2000} with radial velocity between 3,000 km s$^{-1}$ and 
5,000 km s$^{-1}$. Larger circles are the Local Void sample galaxies in
 this study (50 galaxies, see Sect. 2.1). 
 Filled circles are for the 36 galaxies included in the
 final sample (see Sect. 2.3), while open circles are those not in the
 final sample. Dashed lines indicate latitudes in the supergalactic
 coordinates, and the cross shows the position of the North
 supergalactic pole.
 The position opposite to the Local Velocity Anomaly defined by
 \citet{burstein2000} is indicated by a blue asterisk.}
\label{fig:lb}
\end{figure*}

\begin{figure*}
\resizebox{\textwidth}{!}{\includegraphics{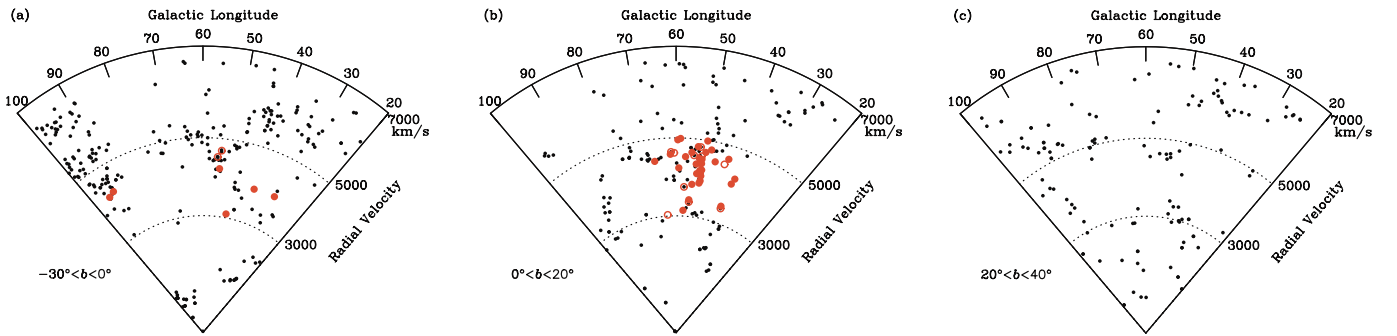}} 
\caption{Distribution of the Local Void sample galaxies in the galactic
 longitude - radial velocity planes. (a), (b) and (c) display the
 galaxies with galactic latitude $-30\degr<b<0\degr$, $0\degr<b<20\degr$ 
 and $20\degr<b<40\degr$, respectively. The meaning of symbols are the
 same as in Fig. \ref{fig:lb}: filled circles represent galaxies in the
 IRAS PSCz catalog \citep{saunders2000} with radial velocities.
 Larger circles are the 50 Local Void sample galaxies in
 this study.
 Larger filled circles are for the 36 galaxies included in the
 final sample, and open circles are those not in the final sample.}
\label{fig:pi}
\end{figure*}

Finally, our observational sample comprises 50 galaxies with measured
redshifts. In Fig. \ref{fig:lb} and \ref{fig:pi} we show the
spatial distribution of these galaxies. It is shown in these figures
that the sample galaxies well represent the population of the far-side
boundary of the Local Void.

\begin{table*}
\centering
\caption{The list of galaxies with new heliocentric radial velocity $V_h$
 measurements in km s$^{-1}$ in the catalog of \citet{roman2000}.}
\begin{tabular}[t]{ccccc}
\hline \hline
CGMW5 ID & Designation & R.A.(2000) & Decl.(2000) & $V_h$ \\
\hline
CGMW5$-$00817 & CGCG 084$-$014 & 18:04:14.7 & +09:20:06 & 6293.5 \\
CGMW5$-$03387 & -- & 18:24:42.9 & +12:20:23 & 5564.3\\
CGMW5$-$05003 & -- & 18:33:28.9 & +22:29:17 & 3864.3\\
CGMW5$-$05171 & -- & 18:34:17.8 & +22:48:27 & 4196.3\\
CGMW5$-$05908 & -- & 18:37:16.1 & +31:46:21 & 4599.3\\
CGMW5$-$06653 & -- & 18:41:19.7 & +24:07:10 & 3968.8\\
CGMW5$-$06881 & -- & 18:42:59.7 & +21:36:18 & 4379.9\\
CGMW5$-$07342 & -- & 18:45:13.5 & +31:57:35 & 7981.9\\
CGMW5$-$10456 & UGC 11417 & 19:14:22.1 & +29:58:34 & 3958.6\tablefootmark{a}\\
\hline
\end{tabular}
\tablefoot{
\tablefoottext{a}{The radial velocity of UGC 11417 has been reported to be 3970
 km s$^{-1}$ by \citet{springob2005}.}
}
\label{tab:optsurvey} 
\end{table*}

\subsection{HI 21-cm line observations}

The 21-cm line observations of the LV galaxies were carried out
with the Nan\c{c}ay radiotelescope. This instrument is a meridian one, with
a half-power beam width at 21-cm of 3.6 arcmin (E-W) $\times$ 22 arcmin
(N-S) at zero declination (nearly the same value for our galaxies,
declinations of which are between $+10\deg$ and $+25\deg$). 
The system temperature is 35K. 
We used a bandwidth of 25 MHz covered by 2048 channels of the
spectrometer, resulting in a velocity resolution of 2.6 km s$^{-1}$. The
observations were performed in an on-off mode, and the integration time
on each galaxy generally ranged from 1 to 2 hours.

\begin{figure*}
\resizebox{\textwidth}{!}{\includegraphics{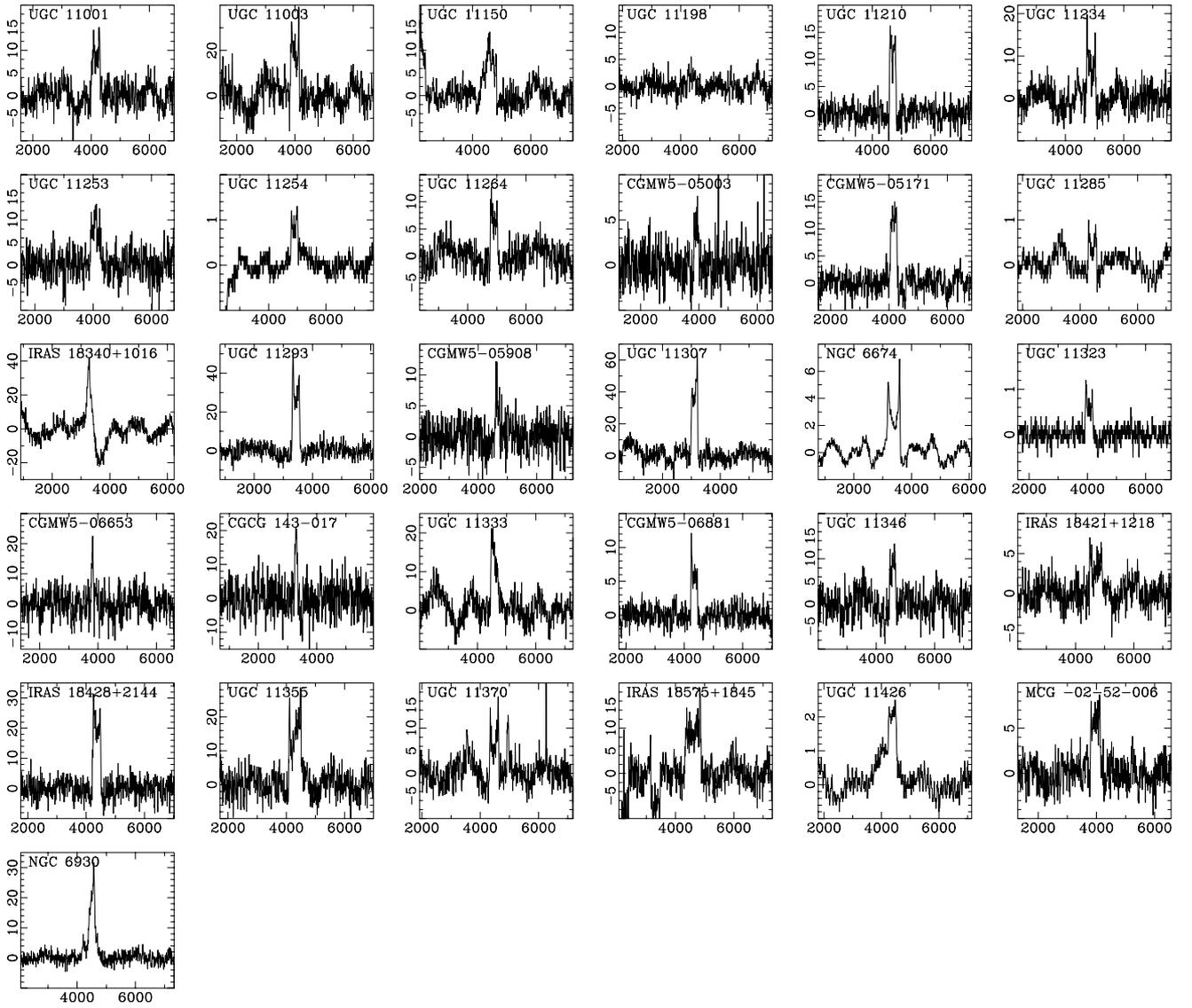}} 
\caption{The HI 21cm line profiles of the 30 Local Void galaxies
 detected with Nan\c{c}ay radiotelescope and of the possibly detected
 one (UGC 11198).
 The horizontal axis is the radial velocity in km s$^{-1}$ and the
 vertical axis is the flux density in mJy. The hanning and boxcar
 smoothing are applied and the subtraction of the polynomial fitted
 baseline is made.} 
\label{fig:nancay_profiles}
\end{figure*}

Forty-three galaxies of our sample were observed between the years 2000
and 2004, leading to 30 detections, one possible detection and 12
non-detections. 
The line profiles of the detected galaxies were reduced using a hanning
and boxcar smoothing, leading to a final velocity resolution of
$2.6\times4=10.4$ km s$^{-1}$. 
The line profiles of the detected galaxies and of the possibly detected
one (UGC 11198) are shown in Fig. \ref{fig:nancay_profiles}, 
after the hanning and boxcar smoothing 
and the subtraction of the polynomial fitted baseline. 
The parameters of interest derived from the profile were obtained,
namely the widths $W_{20}$ and $W_{50}$ of the profile at 20\% and 50\%
of the peak intensity, the heliocentric velocity $V_h$, and
the HI flux $F_H$. All these quantities are given in Table \ref{tab:hi},
with other data for the galaxies that are useful for the present study. 
For five galaxies
(UGC 11254, UGC 11285, CGMW5-05908, UGC 11323, and UGC 11426),
observations were disturbed by the Sun. However, the line widths
can be measured correctly. On the other hand, the profiles of 
IRAS 18340+1016 and NGC 6930 are confused, and their line widths cannot
be measured, so these two galaxies are not listed in Table
\ref{tab:hi}.

The list of the 12 undetected galaxies is the following: 
CGCG 114$-$006, CGCG 172$-$027, CGCG 201$-$043, CGMW5$-$05619, FGC 2187, 
NGC 6586, NGC 6641, NGC 6658, 
UGC 11301, UGC 11353, UGC 11368, and UGC 11369.


Description of Table \ref{tab:hi}:\\
Column (1) Name of the galaxy.\\
(2) Equatorial coordinates $\alpha$, $\delta$ (2000).\\
(3) Galactic coordinates $\ell$, $b$.\\
(4) Heliocentric radial velocity $V_h$ in km s$^{-1}$.\\
(5) Major and minor axes $a_c$ and $b_c$ in arcmin measured at the
isophotal level of 25 mag / arcsec$^2$ in the $B$-band, and corrected
for inclination and for galactic extinction (those data come from
Hyperleda database). \\
(6) Position angle in degrees, from UGC or 2MASS XSC when the galaxy 
is not included in the UGC.\\
(7) Galactic dust attenuation in $V$-band, from the map by
\cite{schlegel1998}.\\
(8) and (9) Widths $W_{20}$ and $W_{50}$ of the HI line in km s$^{-1}$
at 20\% and 50\% of its maximum height, respectively, uncorrected for
velocity resolution, with their uncertainties \citep[after][]{fouque1990}.\\
(10) Measured HI flux $F_H$, in Jy $\times$ km s$^{-1}$, with its
uncertainty:
\begin{equation}
e_{F_H} = 5 F_H^{0.5}(S/N)^{-1} R^{0.5} h^{0.5}
\end{equation}
where S/N is at the point of the profile of maximum intensity, $R$ 
the resolution in km s$^{-1}$, and $h$ the peak intensity of the HI line 
\citep[after][]{fouque1990}. \\ 
(11) HI flux $F_{H,c}$ corrected for beam attenuation; $f_0$ is the
correction factor such that $F_{H,c} = f_0 F_H$. $f_0$ is given by 
\begin{equation}
f_0 = \sqrt{1+\left( \frac{D_{H_{EW}}}{3.6}\right)^2},
\end{equation}
where $D_H$ is the HI diameter of the galaxy, within which half of the
HI mass is contained; $D_{H_{EW}}$ is the projection of this diameter in
the east-west direction, expressed in arcmin. One has $D_H = a_c$
\citep[after][]{hewitt1983}. 
If $\theta$ is the position angle of the galaxy, its east-west corrected
diameter is given by 
\begin{equation}
a_{EW} = \sqrt{a_c^2 sin^2\theta + b_c^2 cos^2\theta}.
\end{equation}
Corrections for beam attenuation are small, only 3\% on an average,
except for NGC 6674 and IRAS 18575+1845, where they reach 20--30\%.\\
(12) Notes.

\subsection{The final sample of galaxies measured in the HI line and selected for use in the IR TF relation}

First we add to our initial sample of 30 detected galaxies 18
other galaxies located on the opposite border of the LV and measured
elsewhere in the HI line (three of them, namely FCG 2187, UGC 11301
and UGC 11369 have not been detected by us). Thus we have a sample of 48
galaxies measured in the HI line at our disposal. 
One can note that 17 among our 30 detected galaxies have also been
detected elsewhere (thus only 13 are newly measured by us).

In order to use the IRTFR in the best conditions, we need to have the
best profile width $W_{20}$. Thus we suppress all the cases of
inaccurate $W$, of possible confusion, and of too narrow a profile
corresponding to dwarf galaxies for which the TFR does not work
correctly. There are 12 such galaxies, namely:
UGC 11150, UGC 11253, UGC 11371, UGC 11552, 
IRAS 18340+1016, NGC 6930 (profile confusion),
CGMW5$-$06653, CGCG 143$-$017, UGC 11333, UGC 11369 (too narrow or
asymmetrical profile), CGMW5$-$05908 ($W_{20}$ not accurate enough), 
and FCG 2187 (uncertain detection).

On the other hand, for galaxies measured by us and elsewhere as well, we
have examined the two profiles obtained. If they were of
equivalent quality, we took the average of the two values for $W_{20}$ 
(after correction for velocity resolution). If one profile was much
better than the other one, its $W_{20}$ value was preferred.

Our final sample comprises 36 galaxies, namely, 19 galaxies for which
our $W_{20}$ or average of ours and others have been taken, four
galaxies measured by us and others and for which the other $W_{20}$ have
been preferred, and 13 galaxies measured only elsewhere.
The final values of $W_{20}$ were corrected for velocity resolution
$R$ by $W_{20}^{cor} = W_{20} - 0.55 R$ 
\citep{bottinelli1990}. 
For our own measurements, the correction is $-4.7$ km s$^{-1}$. These
corrected $W_{20}$ values are used to compute the widths corrected for
inclination and redshift: 
$W_{20}^c = W_{20}^{cor} /  (\mathrm{sin} i (1+z))$ used in IRTFR. 
The $\mathrm{log} W_{20}^c$ values are presented in Table
\ref{tab:nir}.


Description of Table \ref{tab:nir}:\\
(1) Name of the galaxy.\\
(2) Equatorial coordinates $\alpha$, $\delta$ (2000).\\
(3) Galactic coordinates $\ell$, $b$.\\
(4) Heliocentric radial velocity $V_h$ in km s$^{-1}$.\\
(5) Inclination of the galaxy in degrees.\\
(6) $H$-band 20 mag/arcsec$^2$ isophotal elliptical magnitude and its
error.\\
(7) $H$-band magnitude corrected for galactic and internal extinctions.\\
(8) Source of $H$-band photometry. `UH88' and `IRSF' stand for photometry
based on our own observations with the UH 2.2m telescope and the IRSF
1.4m telescope, respectively. `2MAS' stands for data taken from the
2MASS XSC.\\
(9) Logarithm of the width of the HI line at 20\% of its maximum
height, corrected for velocity resolution, inclination, and redshift, 
in km s$^{-1}$.\\
(10) Logarithm of the maximum of the rotation velocity in km s$^{-1}$,
derived from $\log W_c$. See text for details.\\
(11) Source of HI line width data. `Nan\c{c}ay': our own
measurements. `Springob': \citet{springob2005}. `LEDA': the on-line
galaxy database HyperLEDA \citep{paturel2003a}. `Mean': the average
value of measurements by our observations at Nan\c{c}ay, HyperLEDA, and 
\citet{springob2005} (when available).


\subsection{Near-IR ($H$-band) observations and data reduction}

Near-infrared ($H$-band) imaging observations were carried out 
using two facilities. One is the Quick Near-Infrared Camera (QUIRC) 
equipped to the University of Hawaii 2.2m telescope (UH88) at Mauna Kea,
Hawaii, and the other one is the near-infrared Simultaneous Infrared
Imager for Unbiased Survey \citep[SIRIUS:][]{nagashima99,nagayama03} 
on board the Infrared Survey Facility (IRSF) telescope in the South 
African Astronomical Observatory (SAAO) at Sutherland, South Africa. 

Observations with UH88 / QUIRC were carried out in 2001 July 7 -- 9 
and August 4 -- 5 (UT).
The condition was mostly photometric throughout observing dates.
QUIRC has a HAWAII 1024$\times$1024 HgCdTe array with pixel scale of 
$0.19\arcsec$/pixel, yielding a field-of-view of 
$193\arcsec \times 193\arcsec$. 
Total on-source integration times range from 600 sec. to 1200 sec., 
depending on the apparent surface brightness of objects.

Observations with IRSF / SIRIUS were made during 
2003 March 31 -- April 4 and 2004 March 14 -- 22.
Although the SIRIUS camera has a capability of obtaining $J$, $H$, 
and $Ks$ images simultaneously, in the present analysis only $H$-band 
images are used for the $H$-band Tully-Fisher relation.
The field of view and the pixel scale of SIRIUS are 
$7\farcm7 \times 7\farcm7$ and $0\farcs45$, respectively.
Total on-source integration times are 900 sec. 
for UGC 11001 and CGMW5$-$06881, and 1200 sec. for UGC 11003.

Basic data reduction, including dark subtraction, flat-fielding, 
image alignment, and stacking, was done in a standard way using IRAF.
Since the LV region is close to the Galactic plane, 
it is crowded with foreground stars. It is quite important 
to remove these stars before executing the photometry of 
target galaxies, for precise measurement of their apparent 
magnitude. For faint stars we did PSF fitting using Moffat 
profile for each star and subtracted them from reduced images.
For bright stars their profiles are saturated, and we could not 
execute profile fitting. In that case we removed these stars by 
interpolation of counts from surrounding pixels using an IRAF task 
`IMEDIT'.
After the removal of foreground stars, isophotal ellipses with 
$H=20$mag/arcsec$^2$ were defined for 
each galaxy, and we calculated counts within the ellipse.
Photometric zero points were derived for each night using 
near-infrared standard stars. 

For 15 objects in our sample we could not obtain our own 
$H$-band imaging data. For these objects we used data in 
the 2MASS XSC.
We used 20 mag/arcsec$^2$ isophotal elliptical aperture 
magnitude (\verb+h_m_i20e+) in the catalog.
With the galaxies for which we obtained imaging data with UH88 and IRSF, 
we checked the consistency of the photometry between ours and the 2MASS
XSC. For galaxies with relatively fewer foreground stars
in the aperture (which can disturb the automated photometric procedure
adopted in the 2MASS XSC), we found that the difference between our
isophotal magnitudes and those in 2MASS XSC is less than 0.1 mag in
most cases. 
In Table \ref{tab:nir} we list the results of $H$-band photometry for
the final sample galaxies.

\section{Near-infrared Tully-Fisher relation}

We use the Tully-Fisher relation (TFR) to compute the distances of the
LV sample galaxies. We first proceed here to determine the TFR in the 
$H$-band. Generally speaking, 
the Tully-Fisher relation (TFR) is an empirical linear relationship 
between the logarithm of the maximum rotational velocity $V_m$ of any
spiral galaxy and its absolute magnitude $M$, namely,

\begin{equation}
 M = a \log V_m + b,
\label{eq_tfr}
\end{equation}

\noindent
where $a$ and $b$ are constant quantities in a given system of magnitudes 
\citep{tully1977}, and 
$V_m$ can be determined from the width $W$ of the 21-cm HI line or from
the optical rotation curve. This relationship is a powerful and accurate
distance indicator for the spiral galaxies and has been extensively used
for such a purpose \citep[e.g.,][]{sakai2000}.
In the present study it is critically important to 
use near-infrared wavelength photometry data, since the LV region is
close to the Galactic plane, and the effect of Galactic extinction is
significantly reduced in near-infrared wavelengths compared to optical
wavelengths.

\subsection{$H$-band TFR calibration sample}

As a first step, we determine the parameters of the corresponding
IRTFR. For such a purpose, we have to use a sample of spiral galaxies
having known distances, $H$-band magnitudes $m_H$ and rotational velocities
$V_m$ measured in the same systems as those of LV sample galaxies.
As a matter of fact, one can use two possible samples: either a
sample of nearby galaxies having accurate distances measured from
Cepheids or TRGB or a larger sample of more remote galaxies, distances
of which are determined from their redshifts after correction for
attracting various galaxy concentrations. After having tested the two
calibration methods, we concluded that the second one gives more
secure results, mainly due to the large size of the available sample and
because nearby galaxies with large angular dimensions do not have
accurate isophotal $H$-band magnitudes in the 2MASS XSC.
This calibration method does not give the zero point of
the IRTFR, since the absolute magnitudes are computed using the Hubble
law, and thus the zero point depends on the Hubble constant. But this
is convenient for computing the peculiar velocities of the LV galaxies, as
shown in Sect. 4. Hereafter, we use $H_0 = 70$ km s$^{-1}$ Mpc$^{-1}$.

The galaxies of the calibration sample satisfy the same conditions as
those of the LV sample (see Sect. 2.1 and below), and their parameters
used for the calibration have been computed in the same way as those of
the LV sample (see Sect. 4.1 for details of the corrections on radial
velocities). Thus no systemactic difference is introduced between the
two samples.

The parameters needed for the calibration are the $H$-band magnitude
$m_H$, the maximum rotational velocity $V_m$, and the distance $D$ for
each sample galaxy.
Our calibration sample is an all-sky sample of edge-on galaxies with
uniform $H$-band photometry and accurate maximum rotational
velocities. We use the 2MASS XSC for $m_H$ and HyperLEDA 
\citep{paturel2003a} for $V_m$ and recession velocities. 
We put the following conditions on galaxies to be selected for the 
calibration sample.

1. Their recession velocities $V_r$ are lower than 8000 km s$^{-1}$,
allowing accurate correction for infall in nearby clusters.

2. In order to obtain an accurate distance from $V_r$, we only keep 
galaxies having uncertainties on $V_r$ less than 100 km
s$^{-1}$. Moreover, we reject all the nearby galaxies having 
$V_r < 1000$ km s$^{-1}$; indeed, they are generally members of groups,
and the internal motions in groups, about 80 km s$^{-1}$ on the
line-of-sight, introduce an additional scatter on their distances
derived from the redshifts.
For the same reason, we eliminate all the galaxies located within the
two important clusters of galaxies having $V_r < 6000$ km s$^{-1}$, 
namely all the 45 galaxies at less than $15\degr$ and $3\degr$ from the
centers of the Virgo and Coma clusters, respectively.

3. The uncertainties on $V_m$ are lower than 20 km s$^{-1}$; indeed, due
to the high value of the slope of the IRTFR, those uncertainties are the
main source of the observational dispersion of the IRTFR. Thanks to
this condition, the dispersion on the IRTFR due to the measurement
uncertainties is only 0.14 mag. in absolute magnitude, negligible
compared to the intrinsic one ($\sim$ 0.3 mag., see Sect. 3.2).

4. Similar to the LV galaxies, axial ratios are lower than 0.714, and
the morphological types are between 3(Sb) and 8(Sdm).

5. $H$-band magnitudes are taken from the 2MASS XSC. We use 20
mag/sq$''$ isophotal elliptical aperture magnitudes (\verb+h_m_i20e+) as
we did for the LV galaxies.
 $H$-band apparent magnitudes were corrected for
inclination and internal extinction, as well for extinction by our
Galaxy. For internal extinction correction, we first derived the amount
of extinction in $I$-band following \citet{tully1998}:

\begin{equation}
 A_I = \left( 0.92 + 1.63 (\log W_{20}^c - 2.5) \right) \log(1/r),
\label{eq_ai}
\end{equation}

\noindent
where $W_{20}^c$ is the line width at 20\% corrected for inclination in 
km s$^{-1}$ and $r$ is the axis ratio. $A_I$ was converted to $H$-band
extinction by $A_H = 0.5 A_I$ \citep{sakai2000}. 
Amount of galactic extinction toward 
the direction of each sample galaxy was estimated using the extinction
map by \citet{schlegel1998}, and $A_H / E(B-V) = 0.58$ was assumed.

6. In order to correct the IRTFR for the Malmquist bias, we need
our sample to be complete in $m_H$ \citep{theureau2007}. We determine
the completeness limit $m_l$ by plotting $\log[N(\leq m_H)$] versus
$m_H$, where $N(\leq m_H)$ is the number of sample galaxies with an
$H$-band magnitude lower than $m_H$. For an homogeneous distribution of
galaxies, whatever the luminosity function, the completeness to $m_l$ 
is equivalent to the fact that $\log[N(\leq m_H)$] follows the linear
relation: 

\begin{equation}
 \log [N(\leq m_H)] = 0.6 m_H + C,
\label{eq_logN}
\end{equation}

\noindent
for any $m_H \leq m_\mathrm{lim}$, $C$ being a constant.
(For $m_H > m_\mathrm{lim}$, $\log[N(\leq m_H)]$ increases more slowly
than this linear relation.)
For our sample, we obtain $m_\mathrm{lim} = 11.0$. 

Thus our final sample comprises all the galaxies figuring both in 2MASS
and HyperLEDA and satisfying the conditions 1 to 6. The number of
galaxies in the sample is 1463.

\subsection{The Unbiased IRTFR}

Now we proceed to the determination of the unbiased IRTFR, 
following the iterative method devised by \citet{theureau2007}. 
At each iteration, we compute a new IRTFR; the determination of the
IRTFR requires the computation of the absolute $H$-band magnitudes of
the galaxies, which is carried out from their redshifts, corrected for
non-Hubble motions in the same way as those of the galaxies of the LV
sample (see Sect. 4.1). 

In a first step, we determine the IRTFR
from our entire calibration sample, limited however to 
$M_H \leq -22.1$. That cut is made since we found that the slope of the
IRTFR changes at this value, being steeper at $M_H > -22.1$. Moreover,
all the LV galaxies with $m_H \leq 11.0$ are in that part of the IRTFR,
for which the cut does not introduce any classical Malmquist bias. 
The bulk of the calibration galaxies (94\%) remains in the sample 
after this cut. We obtain the best-fit coefficients for the IRTFR: 

\begin{equation}
 M_\mathrm{TF} = -8.06 \log V_m - 5.31 
\label{eq_MTF} 
\end{equation}

\noindent
with $V_m$ in km s$^{-1}$.

\begin{figure*}
\resizebox{\hsize}{!}{\includegraphics{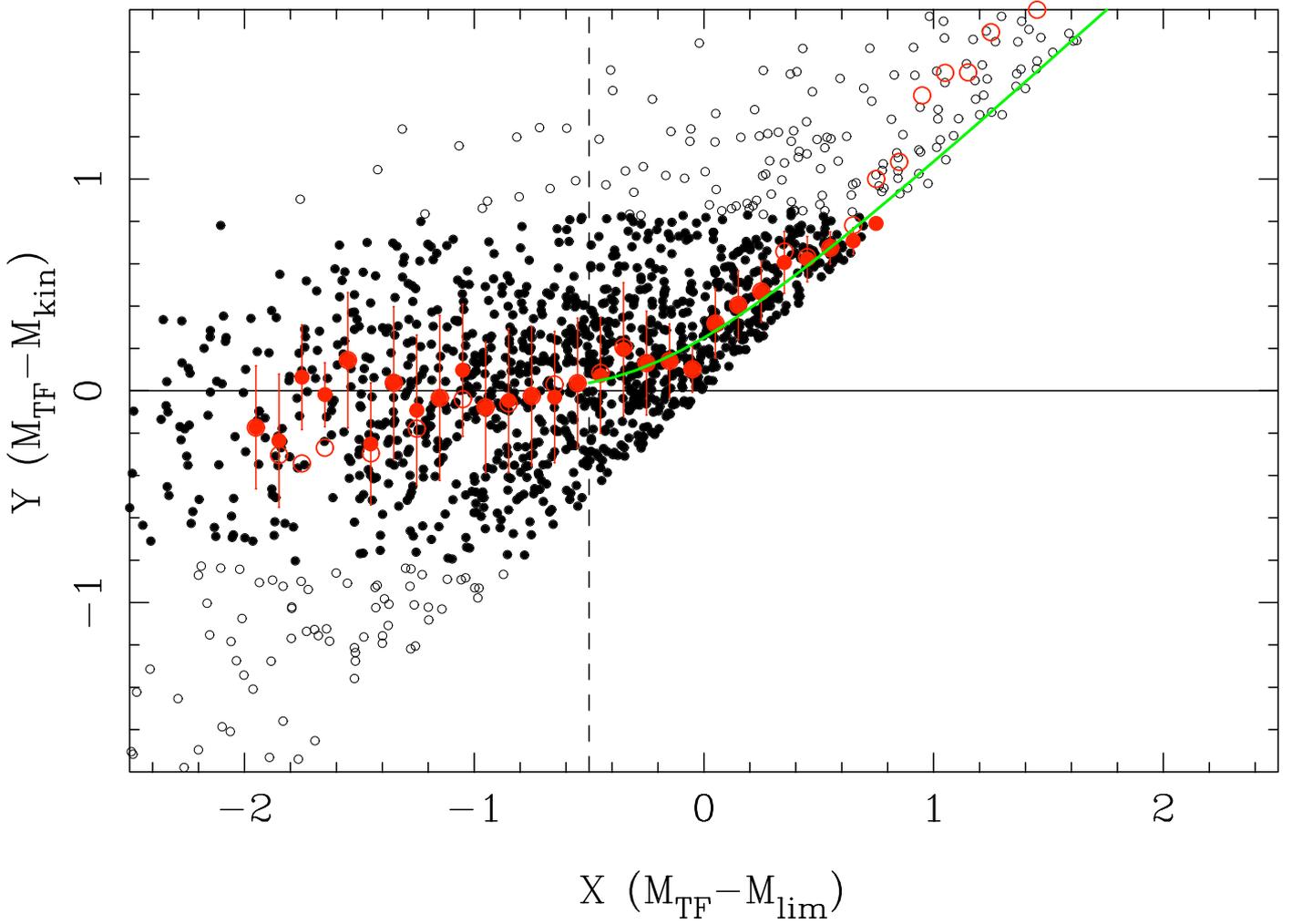}} 
\caption{TFR residuals ($Y = M_\mathrm{TF} - M_\mathrm{kin}$) against the
 normalized distance modulus ($X$) for the calibration sample. Small
 open circles represent the entire sample galaxies, and larger open
 circles show the average values of $Y$ in 0.1 mag. step. Black filled
 circles are galaxies kept after the removal of outliers in the TFR
 plot. Larger filled circles with error bars show the average values for
 them. The vertical dashed line at $X=-0.5$ indicates the upper limit of
 `bias-free' region. The thick green line shows the analytical curve
 giving the expected $\langle Y \rangle$ computed from the dispersion of
 our IRTFR. See text for details.}
\label{fig:tfr_residual}
\end{figure*}

This relation is biased since the sample is not complete in a definite
interval of absolute magnitudes, but only in apparent magnitudes. For a
given sample galaxy, one can determine $M_\mathrm{TF}$ from 
equation (\ref{eq_MTF}),
and also its kinematical absolute $H$-band magnitude $M_\mathrm{kin}$
from the corrected redshift. The quantity 
$Y = M_\mathrm{TF} - M_\mathrm{kin}$ 
exhibits the Malmquist bias, through the uncertainties and the intrinsic
dispersion of the IRTFR, since $\langle Y \rangle > 0$ for our sample
(Fig. \ref{fig:tfr_residual}).

On the other hand, \citet{teerikorpi1975} has shown that the Malmquist
bias depends only on the normalized distance modulus 
$X= M_\mathrm{TF}^c - M_\mathrm{lim}$, where $M_\mathrm{TF}^c$
is the $H$-band absolute magnitude of the galaxy considered as derived
from the unbiased IRTFR, and $M_\mathrm{lim}$ the $H$-band absolute
magnitude cut off corresponding to the limiting apparent magnitude
$m_\mathrm{lim}$. We compute $M_\mathrm{lim}$ using the corrected
redshift of the galaxy; for $M_\mathrm{TF}^c$, 
we take the value of $M_\mathrm{TF}$ obtained from the biased IRTFR 
(equation \ref{eq_MTF}) as a first approximation.

The plot $Y(X)$ is shown in Fig. \ref{fig:tfr_residual}; the absence of
bias for a given value of $X$ is characterized by 
$\langle Y(X) \rangle = 0$. 
One can see that in our sample there is a region free of bias, at
$X \leq -0.5$; for  $X > -0.5$ the bias increases monotonically,
reaching about 1 at $X=1$. 
Thus, in the second iterative step, we keep only the calibration
galaxies having $X\leq -0.5$, in the region free of bias, and 
with those 584 galaxies we obtain

\begin{equation}
 M_\mathrm{TF} = -8.58 \log V_m - 4.06 
\label{eq_MTF2} 
\end{equation}
as a new IRTFR.\footnote{Note that Eq. \ref{eq_MTF2} can be written as
$M_\mathrm{TF} = -8.58 \log V_m - 4.06 + 5 \log (H_0 / 70)$
if we take a different value for $H_0$.}

\begin{figure*}
\resizebox{\textwidth}{!}{\includegraphics{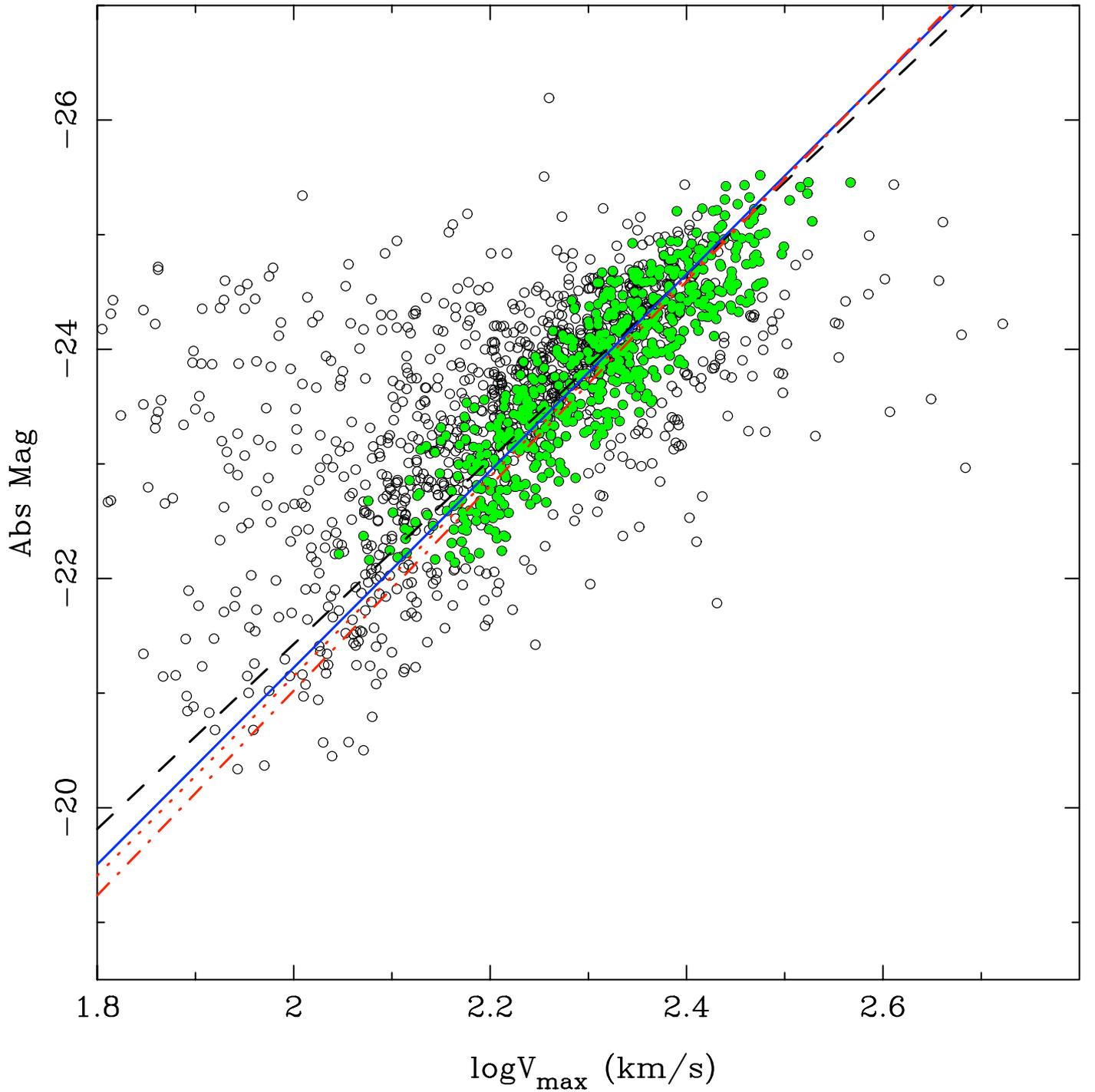}} 
\caption{TFR for the calibration sample. Open circles represent the entire
 calibration sample galaxies, and filled circles are for the `bias-free' 
 subsample, after removal of outliers in the TFR.
 The dashed line is the best-fit for the entire sample 
 (Eq. \ref{eq_MTF}), and the solid line is the best-fit result of the
 first iteration on the unbiased subsample (Eq. \ref{eq_MTF2}). The
 dotted line is the result of the second iteration, which is not
 significantly different from the result of the first iteration. The
 dot-dashed line displays the TFR by \citet{masters2008} after
 conversions in HI line widths and $H$-band magnitude systems 
 (Eq. \ref{eq_TFR1}).
}
\label{fig:tfr_calib}
\end{figure*}

In Fig. \ref{fig:tfr_calib} we show this fit as a solid line over the
TFR distribution of the calibration sample.
However, this IRTFR may not be completely bias free, since the $X$ used
was not equal to $M_\mathrm{TF}^c - M_\mathrm{lim}$, but to
$M_\mathrm{TF}-M_\mathrm{lim}$. So, in a third iterative step, we draw
the new plot ($X$, $Y$) and select the galaxies in the corresponding
region bias-free to compute the next IRTFR. 
We find that the IRTFR does not change significantly.
Thus the unbiased adopted IRTFR is given by equation \ref{eq_MTF2}.  
The corresponding average dispersion of $M_\mathrm{TF}(V_m)$ around the
relation is $\sigma = 0.31$. 
In Fig. \ref{fig:tfr_residual} the data points and average values for the
sample galaxies after the removal of outliers in the TFR plot are shown
as filled circles. Also in Fig. \ref{fig:tfr_residual} the analytic function
of $\langle Y(X) \rangle$ described by \citet{theureau2007} is plotted
for this sample, and it agrees very well with the data. 

Note that \citet{masters2008} have derived an IRTFR in the $H$-band from
a sample of 2MASS calibration galaxies carefully chosen, having total
extrapolated $H$-band magnitudes $M_{H_\mathrm{tot}}$ and accurate HI
profile widths $W_{50}$ at 50\% of the peak intensity. After corrections
for various statistical biases, they obtain the IRTFR in $H$-band as 
$M_{H_\mathrm{tot}}$ versus the width $W_{50}^\mathrm{corr}$ corrected for
inclination. Accounting for the relation between $W_{50}$ and $W_{20}$
\citep{paturel2003b} and the average
difference  $M_{H_\mathrm{tot}} - M_{H_\mathrm{iso}} = -0.15$ between 
total $H$-band magnitudes and our isophotal ones, their relation becomes 

\begin{equation}
  M_{H_\mathrm{iso}} = -8.93 \log V_m - 3.16.
\label{eq_TFR1}
\end{equation}

This relation leads to $M_H$ larger than ours by $0.08 \pm 0.05$
magnitudes in the range $2.2 \leq \log V_m \leq 2.5$ corresponding to
our LV galaxies of interest having $m_H \leq 11.0$. Thus the agreement
is excellent, as that of the scatter of the $M_\mathrm{TF}$ at a given 
$V_m$, which is 0.37 in \citet{masters2008} compared to our value of
0.31.

\section{Non-Hubble residual peculiar velocities of the LV sample galaxies}

The non-Hubble residual peculiar velocity $V_p$ of a galaxy is defined by

\begin{equation}
V_p = V_{corr} - H_0 D
\label{eq_vp}
\end{equation}
where $H_0$ is the Hubble constant, 
$D$ the distance of the galaxy computed here from
the IRTFR (thus independently of the redshift) and $V_{corr}$ is the 
measured radial velocity of the object referred to the centroid of the
Local Group (LG) and corrected for the known local non-Hubble motions
which include different velocities for the LG and the galaxy
considered.
We will use the IRTFR in Eq. \ref{eq_MTF2}. $H_0 D$ does not depend on
the value of $H_0$ adopted if $H_0$ is the same as the one used for the
determination of the IRTFR (in this case 70 km s$^{-1}$ Mpc$^{-1}$).

\subsection{Computation of $V_{corr}$}

To obtain $V_{corr}$, first we convert the measured heliocentric
radial velocities $V_h$ to radial velocities $V_{LG}$ referred to the
centroid of the LG using the equation of \citet{tully2008}:

\begin{equation}
V_{LG} = V_h + 305 \sin l \cos b - 86 \cos l \cos b - 33 \sin b.
\label{eq_vlg}
\end{equation}

Then the known local non-Hubble motions we have to correct $V_{LG}$ for
are the following: 

(1) The repulsion of the LG from the center of the LV, recently
evidenced convincingly by \citet{tully2008} 
thanks to very accurate
distance measurements of galaxies located at less than 10 Mpc from
us, based on the measurements of apparent brightness of TRGB stars. 
The repulsion velocity of the structure at the boundary of the LV (named 
the `Local Sheet', which includes the LG) is reported to be 259 km
s$^{-1}$ toward $\ell=210\degr$, $b=-2\degr$. 
We are looking at a similar repulsion for our sample galaxies. Those
objects and the LG are located nearly at opposite borders of the LV,
and we need to correct the radial velocity of sample galaxies for
this LG motion against the LV. By including the motion of the centroid
of the LG within the Local Sheet 
\citep[66 km s$^{-1}$ towards $\ell=11\degr$, $b=22\degr$;][]{tully2008}
\footnote{There is an error in the calculation of the galactic longitude
of this motion in \citet{tully2008} (i.e., $\ell$ of 
$V_\mathrm{LS}^\mathrm{LG}$) in their Table 3.},
one finds that the centroid of the LG has a velocity of 202 km s$^{-1}$
toward $\ell=215\degr$, $b=5\degr$ with respect to the LV. This results
in corrections of  $\sim -180$ km s$^{-1}$ for radial velocities of our
sample galaxies. 

(2) The infalls towards three nearby mass concentrations, namely the
Virgo cluster, the Great Attractor (GA) and the Shapley
supercluster. Such corrections are necessary since the infall velocities
are quite different for the sample galaxies and for the LG.
The infall corrections have been carried out following
\citet{mould2000}. 
In brief, those authors use a simple multi-attractor model; they assume
the flows to be independent, thus the respective velocity infalls add
to each other. The infall velocity $V_f$ towards each attractor at the
level of the LG is known. If one assumes that the attractor has the
spherical symmetry with a density profile $\rho(r) \propto r^{-\gamma}$,
then the infall velocity $V(r)$ is: $V(r) \propto  r^{1-\gamma}$, and
one can compute the projected velocity component $V_{inf}$ on the
line-of-sight of the galaxy oriented towards the galaxy, of the infall
velocity of the object caused by the attractor, as seen from the
infalling LG, namely,

\begin{equation}
V_{inf} = V_f (\frac{r_a \cos \theta - r_0}{r_{0a}}) 
(\frac{r_{0a}}{r_a})^{1-\gamma} - V_f \cos \theta,
\label{eq_vinf} 
\end{equation}
where $\theta$ is the angle between the directions of the galaxy and of
the attractor as seen from us, $r_0$ is the distance of the object,
$r_a$ is the distance of the attractor,
and $r_{0a}$ is the distance between the galaxy and the attractor:
$r_{0a} = \sqrt{r_0^2 + r_a^2 - 2 r_0 r_a cos \theta}$.

Following \citet{mould2000}, 
we take $\gamma = 2$, which fits the Virgo cluster. Note that in
equation (\ref{eq_vinf}), the two successive terms represent the
difference of the respective projections, on the line-of-sight of the
galaxy, of the object and of the LG infall velocities towards the
attractor. 

Coordinates of the attractors and $V_f$ values are taken from 
\citet{mould2000}; we have adopted for $r_a$: 17.4 Mpc, 65.8 Mpc and
190.3 Mpc for Virgo, GA and Shapley supercluster, respectively,
corresponding to $H_0 = 70$ km s$^{-1}$ Mpc$^{-1}$ for velocities of the
attractors corrected for infalls. The distances $r_0$ of the LV galaxies
have been computed from their redshifts referred to the centroid of the
LG.

The values of $V_{inf}$ for our galaxies are on the order of 
50 km s$^{-1}$ Mpc$^{-1}$, 200 kms s$^{-1}$ Mpc$^{-1}$, and 
20 kms s$^{-1}$ Mpc$^{-1}$ for Virgo, GA, and Shapley infalls,
respectively. Taking the repulsion of the LG from the
center of the LV into account, the total corrections to $V_{LG}$, 
$\delta V_{LG} = \delta V_{LV} - V_{inf}$ are in fact quite small, about
a few tens of km s$^{-1}$.
If the infall corrections for these nearby mass concentrations are
not applied, the peculiar velocities for most of the galaxies in the
final sample are $\sim$100 -- 300 km s$^{-1}$ smaller than when using
these corrections. Those velocities are near the expected LV 
expansion, thus infall corrections have to be applied here.

The corrected velocities are given in Table \ref{tab:vpec}.

\subsection{Distances of the LV galaxies}

We compute the distances of the LV galaxies using the IRTFR corrected
for Malmquist bias (determined in Sect. 3.2). 
In order not to introduce any other bias, we
have to treat our LV sample exactly as the calibration sample, in 
particular to limit it to $m_H \leq 11.0$ and $M_H \leq -22.1$, and also
compute $V_m$ the same way.
Only 15 galaxies among the 36 of our LV sample have 
$m_H \leq 11.0$, and it happens that all those 15 are in the part 
$\log V_m \geq 2.10$, which corresponds to $M_H \leq -22.1$.

\subsubsection{Determination of the maximum rotational velocity $V_m$}

The HyperLEDA extragalactic database provides the maximum rotational
velocity $V_m$ for a number of galaxies. We have used these $V_m$ values
for those eight galaxies in our LV sample for which we took HyperLEDA
21-cm data. For the other LV galaxies, $V_m$ was determined using the
tight correlation between $V_m$ and $W_{20}^c$ obtained from the
calibration galaxies, namely,

\begin{equation}
 \log V_m = 1.12 \log W_{20}^c - 0.67,
\end{equation}
where $W_{20}^c$ is obtained from the HI profile width at 20\% 
of the peak value corrected for resolution by

\begin{equation}
 W_{20}^c = W_{20} / \sin i (1+z),
\end{equation}
where $i$ is the inclination of the galaxy and $z$ the redshift. 
Taking the apparent axis values obtained with super-coadded image in
2MASS using $J$, $H$ and $K$-bands, we derive $i$ from the apparent 
axis ratio $r$ by

\begin{equation}
 \cos i = \sqrt{\frac{r^2-r_0^2}{1-r_0^2}},
\end{equation}
where $r_0$ is the intrinsic minor to major axis ratio for spiral
galaxies. Following many previous studies on TFR
\citep[e.g.,][]{tully1977,sakai2000,masters2008}, we take $r_0 = 0.2$.

\subsubsection{Computation of the distances and correction for Malmquist bias}

The distance modulus $\mu$ of any galaxy of our LV sample is given by
\begin{equation}
 \mu = m_H^c - M_H^c,
\end{equation}
where $m_H^c$ and $M_H^c$ are the apparent and absolute $H$-band
magnitudes corrected for galactic and internal extinction, respectively.
In Table \ref{tab:vpec} we show the distances for the 15 galaxies with 
$m_H < 11.0$. The photometric data are either from our own observations 
or from 2MASS XSC when our data are not available, as shown in Table 
\ref{tab:nir}. We also show the case where only 2MASS XSC photometry 
is used. Using only 2MASS photometry eliminates a possibility of 
systematic difference between data based on different facilities,
but since the LV is located at low Galactic latitudes, 
the 2MASS photometry would suffer from contamination by 
foreground stars. In our own photometry we carefully removed 
foreground stars (Sect. 2.4), so this effect should be alleviated.

As explained above (Sect. 3.2), $M_H^c$ is determined from the value
of the absolute magnitude $M_\mathrm{TF}(V_m)$ given by the IRTFR free
of Malmquist bias.
However, there is a bias in the LV sample since it is limited in
apparent magnitude. 
For a given $V_m$, the less luminous galaxies are not
included in the sample, and this causes the average absolute magnitude
$\langle M \rangle$ at $V_m$ to be biased toward being more luminous
than the true value. This effect is shown in Fig. \ref{fig:tfr_residual}
for the case of the calibration sample. In this figure, 
$X$ is a normalized distance modulus $M_\mathrm{TF} - M_\mathrm{lim}$ where 
$M_\mathrm{lim} = m_\mathrm{lim} - \mu_\mathrm{kin}$ is an absolute
magnitude cut off, $\mu_\mathrm{kin}$ the distance modulus
computed from the radial velocity, and 
$Y$ a normalized magnitude $M_\mathrm{TF}-M_\mathrm{kin}$, where 
$M_\mathrm{kin}$ is the absolute magnitude corresponding to $m_H$: 
$M_\mathrm{kin} = m_H - \mu_\mathrm{kin}$. As discussed in Sect. 3.2,
there is a bias in $\langle Y \rangle$ at $X > -0.5$, and thus in order
to use galaxies with $X > -0.5$ to derive the average peculiar velocity, 
we need to correct the bias by applying the correction to the
absolute magnitude obtained from IRTFR:
$M_H^c(X)=M_\mathrm{TF}(V_m) - \langle Y(X) \rangle$.
The value of $\langle Y(X) \rangle$ is computed analytically from the TF
dispersion, following \citet{theureau2007}.

Among the 15 galaxies with $m_H \leq 11.0$, there are six galaxies 
with $X > -0.5$. Correction factors for the distances are less than 
8\%, except for UGC 11426 where it is 26\%.

\subsection{Residual peculiar velocities of the LV galaxies}

\begin{figure*}
\resizebox{\hsize}{!}{\includegraphics{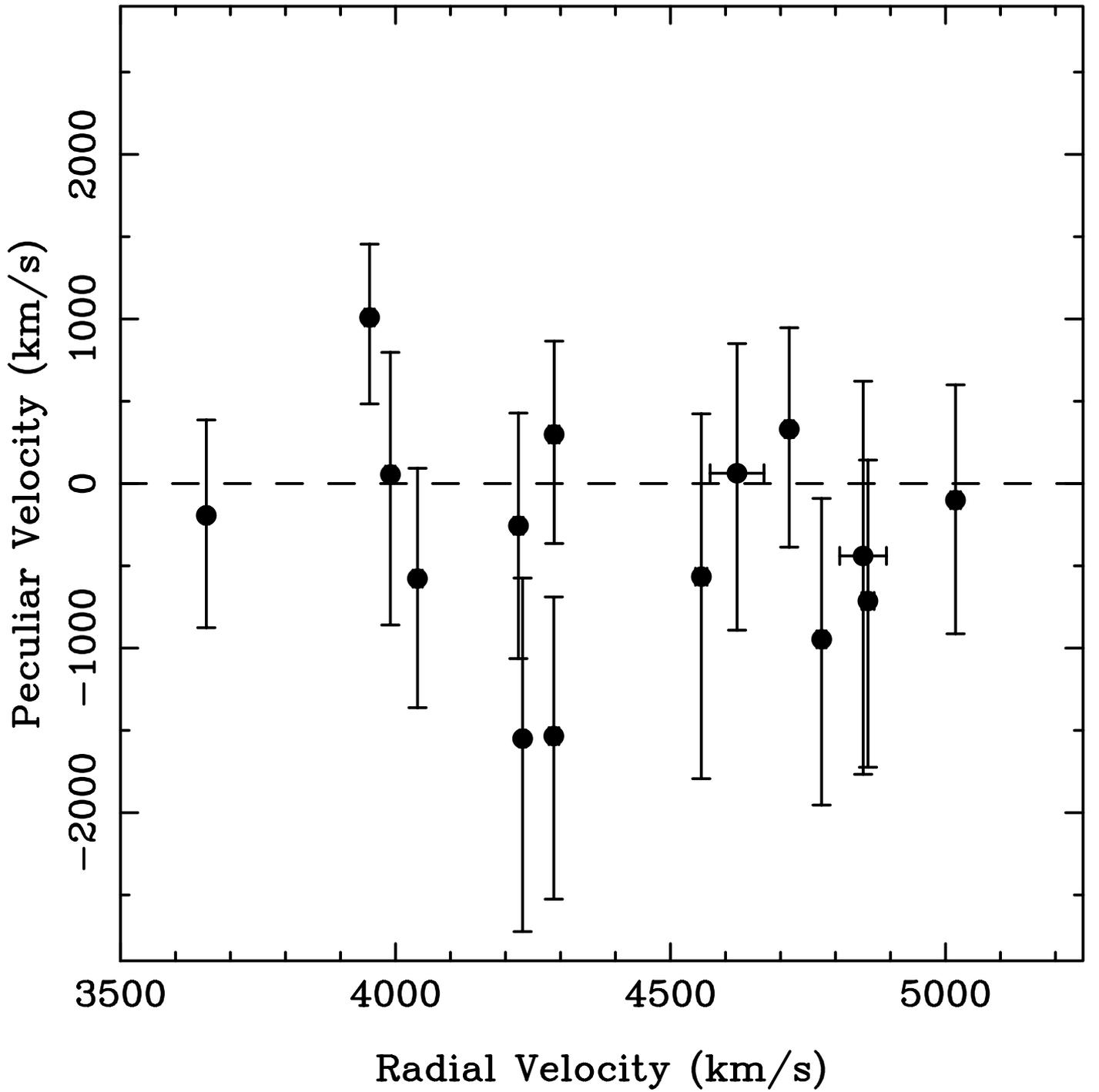}} 
\caption{Corrected radial velocities and peculiar velocities of the
 `unbiased' Local Void sample galaxies. For galaxies with the normalized
 distance modulus $X=M_\mathrm{TF}-M_\mathrm{lim} > -0.5$, the bias
 corrections were applied (see text for details).}
\label{fig:lvpecvel}
\end{figure*}

We compute the residual peculiar velocities $V_p$ with Eq. \ref{eq_vp}, 
using the distance $D$ determined in the previous section.
`Bias-corrected' distances are used for the galaxies with $X>-0.5$.
The calculated $V_p$ are shown in Table \ref{tab:vpec}.
In the results based on UH88 / IRSF / 2MASS photometry, 10 galaxies have
negative values of $V_p$, five have positive ones.
In Fig. \ref{fig:lvpecvel} we show the distribution of $V_p$ 
against the radial velocities 
$V_\mathrm{corr}$. There appears to be no significant correlation
between the radial velocities and the peculiar velocities.

Accounting for the uncertainties on each $V_p$ as listed in Table 
\ref{tab:vpec}, we obtain the average value of $V_p$:
$\langle V_p \rangle$ is $-419 +208 -251$ km s$^{-1}$ 
for the case with UH88 / IRSF / 2MASS photometry, and 
$\langle V_p \rangle$ is $-319 +204 -246$ km s$^{-1}$ 
for the case with 2MASS XSC photometry alone.
Thus $\langle V_p \rangle$ is not significantly different from a zero
value. The values within a 3$\sigma$ error range 
from $-1172$ km s$^{-1}$ to 205 km s$^{-1}$
for UH88 / IRSF / 2MASS photometry.
If the LV has a general expansion, one may think that the 
expansion at the far boundary of the LV we investigate 
is comparable to that of the Local Sheet, i.e., 
259 km s$^{-1}$ \citep{tully2008}, and it is 
slightly out of but very close to 
the 3$\sigma$ error range of our results. 
Due to the size of the error, we cannot evidence such an expansion of
the far-side boundary of the LV.
The size of the errors in our peculiar velocities comes mainly from the
intrinsic scatter of the IRTFR ($\sigma = 0.31$ in absolute
magnitude). There are some other sources, such as the uncertainties on
the widths of the HI line profiles, uncertainty in $H$-band photometry, 
and in radial velocities, but they are less than the effect of the 
scatter of the TFR.

One can note that the dispersion of the $V_p$ around their average is
also about 200 km s$^{-1}$, as for the expected uncertainty, which
shows that there is no systematic variations in $V_p$ among our sample
galaxies. 

\section{Summary and conclusions}

The LV, the nearest void of galaxy distribution from us, 
is expected to undergo a general expansion, as in any
void of galaxies, due to the lack of matter within it. \citet{tully2008}
show that the Local Group and galaxies near the Local Group, i.e.,
the Local Sheet at the edge of the LV, move away from
the center of the LV with a velocity of 259 km s$^{-1}$.

In the present study, we investigated the peculiar velocities of the
galaxies located at the opposite edge of the LV with respect to the
Local Group to see if they show an expansion from the center of the
LV, by using the IR Tully-Fisher relation to compute their distances. 
The sample galaxies have an edge-on spiral morphology and radial
velocities between 3000 km s$^{-1}$ and 5000 km s$^{-1}$. 
Gathering 19 HI line width
measurements by ourselves and those in the literature leads to a final 
number of sample galaxies of 36. We also made $H$-band photometry for
the majority of the sample galaxies. To derive the IR
Tully-Fisher relation, 
we used a large sample of galaxies having maximum rotational velocities 
in HyperLEDA \citep{paturel2003a} and $H$-band isophotal magnitudes 
in 2MASS XSC, complete to $m_H=11.0$. The IRTFR free from the Malmquist
bias was obtained from that sample, and then was used to compute the 
distances of the 15 LV galaxies having $m_H \leq 11.0$. After the
corrections for the infall motions toward the nearby clusters /
concentrations and the motion of the Local Group away from the center of
the LV, the residual peculiar velocities $V_p$ for the 15 LV galaxies
have been obtained. The average value after the correction for the
Malmquist bias (which is thought to affect $V_p$ of some of the
sample galaxies) is $\langle V_p \rangle = -419^{+208}_{-251}$ 
km s$^{-1}$. This is not significantly different from zero, and it does
not reject the possibility that these galaxies have a motion against the
LV equivalent to that of the Local Sheet (259 km s$^{-1}$).

\citet{padilla2005} made a $\Lambda$ CDM numerical simulation on the
properties of dark matter halos and galaxies around voids and find a
linear relation between the maximum outflow velocity $v_\mathrm{max}$
and the distance from the center of the void $r_\mathrm{void}$: 
$v_\mathrm{max} = v_0 r_\mathrm{void}$, where the best-fit value of
$v_0$ is 14.5 km s$^{-1}$ $h$ Mpc$^{-1}$. If we adopt the radius of the
void to be 2500 km s$^{-1}$ and $h=0.7$, this gives the $v_\mathrm{max}$
of 360 km s$^{-1}$.
Such a value is significantly higher than the motion of 259$\pm$25 
km s$^{-1}$ of the Local Sheet away from the LV, and at 3.7$\sigma$ from 
our $\langle V_p^c \rangle$. Thus it does not seem to account correctly
for the expansion of the LV. However, the geometry of the LV
is more complex than a simple sphere, consisting in a void within two
large voids \citep{tully2008}. Thus the maximum velocity
\citet{padilla2005} measured in their simulated voids might not be fully
appropriate for the comparison with the expansion velocity of the LV.

Finally, the uncertainty $\sim$200 km s$^{-1}$ of our 
$\langle V_p^c \rangle$ is not sufficient for proving the expansion
found by \citet{tully2008} for the Local Sheet.
Smaller uncertainty would be achieved if we were able to use more
galaxies -- i.e., using galaxies with apparent magnitude fainter than
$m_H = 11.0$. However, since the number of galaxies at the far-side of
the LV, which can be used for IRTFR, would not exceed $\sim$50, 
we may need an alternative, more accurate distance estimator to 
conclusively know whether the opposite edge of the LV undergoes an
expansion.

\begin{acknowledgements}
We thank staff members of the telescope facilities used in this work
 (Okayama Astrophysical Observatory, Nan\c{c}ay radiotelescope, the
 Infrared Survey Facility, and the University of Hawaii 2.2m telescope)
 for their support during observations. We would like to thank the
 referee (B. Tully) for helpful comments that improved the paper.
 II is grateful to M. Sait\={o} who raised the initial idea of this
 work, A. T. Roman for his participation in the
 observations in Okayama, and K. Nakanishi and K. Ohta for supporting
 observations and giving thoughtful suggestions. II was supported by a
 Research Fellowship of the Japan Society for the Promotion of Science
 (JSPS) for Young Scientists during parts of this research.
\end{acknowledgements}


\bibliographystyle{aa}
\bibliography{16687.bib}



\longtab{2}{
\begin{landscape}
\begin{table*}
\caption{Summary of the results of our HI 21 cm line observations with
 the Nan\c{c}ay radiotelescope.}
\begin{tabular}{rccccccccccl}
\hline \hline
\multicolumn{1}{c}{Name} & $\alpha$, $\delta$ (2000) & $l$, $b$ & $v_h$ & Size & P.A. & $A_V$ & $W_{20}$ &  $W_{50}$ & HI flux & $F_{H,c}$ & comment \\
\multicolumn{1}{c}{(1)} &  (2) & (3) & (4) & (5) & (6) & (7) & (8) & (9) & (10)  & (11) & (12) \\
\hline
      UGC 11001 & 17:50:13.6  ~ +14:17:12 &  39.2207 ~   + 19.8857 & 4216 & 1.5 $\times$ 0.7 & 133 & 0.296 & 307 $\pm$ 16 & 292 $\pm$ 10 & 2.78 $\pm$ 0.27 & 2.95 &   \\
      UGC 11003 & 17:50:41.9  ~ +14:49:18 &  39.7894 ~   + 19.9999 & 4056 & 1.1 $\times$ 0.3 & 165 & 0.286 & 304 $\pm$ 16 & 286 $\pm$ 10 & 6.01 $\pm$ 0.57 & 6.04 &   \\
      UGC 11150 & 18:12:06.9  ~ +25:35:43 &  52.4097 ~   + 19.5129 & 4808 & 1.4 $\times$ 0.3 &   7 & 0.487 & 314 $\pm$ 30 & 283 $\pm$ 19 & 3.77 $\pm$ 0.28 & 3.78 &  Confusion by a companion \\
      UGC 11210 & 18:22:07.7  ~ +21:10:38 &  49.0520 ~   + 15.6971 & 4776 & 1.3 $\times$ 0.1 & 131 & 0.479 & 236 $\pm$  9 & 219 $\pm$  6 & 2.69 $\pm$ 0.17 & 2.79 &  \\ 
      UGC 11234 & 18:26:32.8  ~ +22:25:12 &  50.6586 ~   + 15.2531 & 4987 & 1.0 $\times$ 0.4 & 132 & 0.467 & 308 $\pm$ 13 & 297 $\pm$  8 & 2.72 $\pm$ 0.26 & 2.79 &   \\
      UGC 11253 & 18:29:39.7  ~ +23:04:06 &  51.5710 ~   + 14.8537 & 4145 & 1.0 $\times$ 0.2 & 111 & 0.356 & 332 $\pm$ 16 & 316 $\pm$ 10 & 2.90 $\pm$ 0.30 & 3.00 &  Confused profile \\
      UGC 11254 & 18:29:42.9  ~ +30:26:14 &  58.6622 ~   + 17.6787 & 4987 & 1.3 $\times$ 0.2 & 105 & 0.387 & 256 $\pm$ 12 & 242 $\pm$  7 & 0.23 $\pm$ 0.02 & 0.24 &  Profile disturbed by the Sun \\
      UGC 11264 & 18:31:46.0  ~ +31:05:23 &  59.4700 ~   + 17.5052 & 4951 & 1.2 $\times$ 0.5 &  55 & 0.366 & 264 $\pm$ 15 & 251 $\pm$ 10 & 1.85 $\pm$ 0.21 & 1.94 &   \\
    CGMW5-05003 & 18:33:29.0  ~ +22:29:16 &  51.3891 ~   + 13.8136 & 3864 & 0.5 $\times$ 0.2 & 130 & 0.430 & 175 $\pm$ 47 & 142 $\pm$ 30 & 0.73 $\pm$ 0.14 & 0.73 &   \\
    CGMW5-05171 & 18:34:17.8  ~ +22:48:27 &  51.7685 ~   + 13.7717 & 4196 & 1.1 $\times$ 0.4 & 166 & 0.416 & 236 $\pm$  7 & 224 $\pm$  4 & 2.78 $\pm$ 0.18 & 2.79 &   \\
      UGC 11285 & 18:35:14.2  ~ +22:29:57 &  51.5706 ~   + 13.4480 & 4500 & 1.4 $\times$ 0.5 &  42 & 0.487 & 305 $\pm$ 16 & 295 $\pm$ 10 & 0.14 $\pm$ 0.02 & 0.15 &  Profile disturbed by the Sun \\
      UGC 11293 & 18:36:41.7  ~ +10:26:19 &  40.6165 ~   +  7.9839 & 3443 & 1.0 $\times$ 0.3 &  80 & 0.969 & 270 $\pm$  9 & 249 $\pm$  6 & 8.11 $\pm$ 0.29 & 8.40 &   \\
    CGMW5-05908 & 18:37:16.1  ~ +31:46:21 &  60.5889 ~   + 16.6691 & 4599 & 0.7 $\times$ 0.3 &  92 & 0.338 &  92 $\pm$ 40 &  51 $\pm$ 25 & 0.93 $\pm$ 0.16 & 0.95 &  Profile disturbed by the Sun \\ 
      UGC 11307 & 18:37:55.3  ~ +27:47:36 &  56.8255 ~   + 15.0121 & 3145 & 1.3 $\times$ 0.5 &  50 & 0.406 & 241 $\pm$ 11 & 217 $\pm$  7 & 9.53 $\pm$ 0.43 & 9.93 &   \\
       NGC 6674 & 18:38:33.9  ~ +25:22:30 &  54.5925 ~   + 13.9190 & 3428 & 4.3 $\times$ 2.2 & 143 & 0.330 & 470 $\pm$ 25 & 439 $\pm$ 16 & 1.50 $\pm$ 0.07 & 1.89 &   \\
      UGC 11323 & 18:40:57.9  ~ +23:05:18 &  52.6770 ~   + 12.4885 & 4116 & 0.9 $\times$ 0.4 &  85 & 0.444 & 279 $\pm$ 27 & 251 $\pm$ 17 & 0.18 $\pm$ 0.01 & 0.19 &  Profile disturbed by the Sun \\
    CGMW5-06653 & 18:41:19.7  ~ +24:07:09 &  53.6747 ~   + 12.8351 & 3968 & 0.9 $\times$ 0.3 & 103 & 0.316 &  89 $\pm$ 51 &  56 $\pm$ 33 & 1.15 $\pm$ 0.14 & 1.18 &  Narrow profile(dwarf?) \\
   CGCG 143-017 & 18:42:15.1  ~ +24:53:51 &  54.4915 ~   + 12.9624 & 3372 & 1.0 $\times$ 0.3 & 100 & 0.459 & 115 $\pm$  8 & 106 $\pm$  5 & 1.75 $\pm$ 0.31 & 1.82 &  Narrow profile \\
      UGC 11333 & 18:42:26.2  ~ +32:22:31 &  61.6013 ~   + 15.8829 & 4644 & 1.0 $\times$ 0.4 & 151 & 0.246 & 186 $\pm$ 51 & 134 $\pm$ 33 & 2.48 $\pm$ 0.28 & 2.52 &  Asymmetrical profile \\
    CGMW5-06881 & 18:42:59.7  ~ +21:36:18 &  51.4990 ~   + 11.4415 & 4380 & 0.9 $\times$ 0.3 &  38 & 0.558 & 241 $\pm$ 10 & 228 $\pm$  6 & 1.54 $\pm$ 0.11 & 1.56 &   \\
      UGC 11346 & 18:44:16.8  ~ +25:15:23 &  55.0207 ~   + 12.6905 & 4645 & 1.4 $\times$ 0.6 &  77 & 0.589 & 226 $\pm$ 21 & 209 $\pm$ 13 & 1.85 $\pm$ 0.27 & 1.98 &   \\
IRAS 18421+1218 & 18:44:31.2  ~ +12:21:52 &  43.2150 ~   +  7.1160 & 4649 & 1.6 $\times$ 0.4 & 160 & 1.418 & 449 $\pm$ 33 & 427 $\pm$ 21 & 1.47 $\pm$ 0.17 & 1.49 &  \\ 
IRAS 18428+2144 & 18:44:57.7  ~ +21:47:26 &  51.8669 ~   + 11.1054 & 4389 & 0.8 $\times$ 0.3 & 150 & 0.722 & 301 $\pm$ 16 & 273 $\pm$ 10 & 6.08 $\pm$ 0.32 & 6.12 &   \\
      UGC 11355 & 18:47:57.0  ~ +22:56:33 &  53.2278 ~   + 10.9655 & 4360 & 1.7 $\times$ 0.6 & 126 & 0.709 & 456 $\pm$ 31 & 421 $\pm$ 20 & 7.37 $\pm$ 0.44 & 7.89 &   \\
      UGC 11370 & 18:51:43.5  ~ +26:33:17 &  56.9387 ~   + 11.7035 & 4525 & 1.5 $\times$ 0.5 &  48 & 0.703 & 317 $\pm$ 29 & 297 $\pm$ 18 & 2.36 $\pm$ 0.23 & 2.48 &   \\
IRAS 18575+1845 & 18:59:47.2  ~ +18:49:49 &  50.6935 ~   +  6.6900 & 4689 & 2.8 $\times$ 0.7 &  65 & 1.892 & 571 $\pm$ 57 & 522 $\pm$ 36 & 5.25 $\pm$ 0.33 & 6.46 &   \\
      UGC 11426 & 19:18:21.7  ~ +34:50:13 &  67.0505 ~   + 10.0705 & 4430 & 1.0 $\times$ 0.7 &  55 & 0.465 & 342 $\pm$ 32 & 302 $\pm$ 20 & 0.68 $\pm$ 0.05 & 0.71 &  Profile disturbed by the Sun \\
 MCG -02-52-006 & 20:29:35.1 ~$-$12:30:12 &  32.4273 ~ $-$ 27.3580 & 3932 & 1.0 $\times$ 0.4 & 110 & 0.123 & 356 $\pm$ 22 & 333 $\pm$ 14 & 2.01 $\pm$ 0.17 & 2.09 &   \\
\hline
\end{tabular}
\label{tab:hi}
\end{table*}
\end{landscape}

}

\longtab{3}{
\begin{landscape}
\begin{table*}
\centering
\caption{Summary of the near-infrared photometry and corrected HI line
 widths for the final Local void sample galaxies.}
\begin{tabular}{rcccccccccc}
\hline \hline
\multicolumn{1}{c}{Name} & $\alpha$, $\delta$ (2000) & $l$, $b$   & $v_h$ & incl. & $m_H$ & $m_c$ & NIR Src & $\log W_c$ & $\log V_{\mathrm max}$ & HI Src  \\
\multicolumn{1}{c}{(1)}  & (2)                       & (3)        & (4)   & (5)   & (6)   & (7)   & (8)     & (9)        & (10)                   & (11) \\
\hline
      UGC 11001 & 17:50:13.6  ~ +14:17:12 &  39.2207 ~   +19.8857 & 4216 & 69.30 & $11.82\pm0.04$ & $11.57\pm0.04$ & IRSF & $2.503\pm0.024$ & $2.134\pm0.027$ &     Mean \\
      UGC 11003 & 17:50:41.9  ~ +14:49:18 &  39.7894 ~   +19.9999 & 4065 & 84.63 & $11.83\pm0.04$ & $11.39\pm0.04$ & IRSF & $2.512\pm0.023$ & $2.144\pm0.026$ &     Mean \\
      UGC 11142 & 18:11:38.6  ~ +25:39:27 &  52.4290 ~   +19.6352 & 4511 & 78.46 & $11.46\pm0.02$ & $11.03\pm0.07$ & UH88 & $2.589\pm0.022$ & $2.210\pm0.011$ &     LEDA \\
      UGC 11210 & 18:22:07.7  ~ +21:10:38 &  49.0520 ~   +15.6971 & 4776 & 76.81 & $14.35\pm0.03$ & $13.96\pm0.05$ & UH88 & $2.369\pm0.023$ & $1.984\pm0.026$ &   Nan\c{c}ay \\
      UGC 11234 & 18:26:32.8  ~ +22:25:12 &  50.6586 ~   +15.2531 & 4987 & 62.11 & $11.42\pm0.03$ & $11.18\pm0.03$ & UH88 & $2.528\pm0.025$ & $2.162\pm0.028$ &   Nan\c{c}ay \\
      UGC 11246 & 18:28:23.9  ~ +22:44:12 &  51.1353 ~   +14.9879 & 4071 & 69.30 & $10.16\pm0.03$ & $ 9.89\pm0.03$ & UH88 & $2.746\pm0.013$ & $2.406\pm0.015$ & Springob \\
      UGC 11254 & 18:29:42.9  ~ +30:26:14 &  58.6622 ~   +17.6787 & 4987 & 60.67 & $13.52\pm0.05$ & $13.30\pm0.05$ & 2MAS & $2.452\pm0.026$ & $2.077\pm0.029$ &   Nan\c{c}ay \\
      UGC 11261 & 18:31:06.7  ~ +22:24:36 &  51.0866 ~   +14.2829 & 3945 & 67.86 & $12.99\pm0.03$ & $12.73\pm0.03$ & UH88 & $2.372\pm0.031$ & $2.070\pm0.026$ &     LEDA \\
      UGC 11264 & 18:31:46.0  ~ +31:05:23 &  59.4700 ~   +17.5052 & 4951 & 62.11 & $12.24\pm0.05$ & $12.02\pm0.05$ & 2MAS & $2.459\pm0.032$ & $2.085\pm0.036$ &   Nan\c{c}ay \\
    CGMW5-05003 & 18:33:29.0  ~ +22:29:16 &  51.3891 ~   +13.8136 & 3864 & 57.73 & $13.03\pm0.03$ & $12.81\pm0.03$ & UH88 & $2.298\pm0.130$ & $1.905\pm0.145$ &   Nan\c{c}ay \\
    CGMW5-05171 & 18:34:17.8  ~ +22:48:27 &  51.7685 ~   +13.7717 & 4196 & 72.21 & $13.26\pm0.09$ & $12.96\pm0.09$ & 2MAS & $2.379\pm0.019$ & $1.996\pm0.021$ &   Nan\c{c}ay \\
      UGC 11285 & 18:35:14.2  ~ +22:29:57 &  51.5706 ~   +13.4480 & 4500 & 75.23 & $11.78\pm0.02$ & $11.42\pm0.04$ & UH88 & $2.502\pm0.023$ & $2.133\pm0.026$ &     Mean \\
      UGC 11293 & 18:36:41.7  ~ +10:26:19 &  40.6165 ~   + 7.9839 & 3443 & 60.67 & $11.68\pm0.02$ & $11.35\pm0.03$ & UH88 & $2.478\pm0.021$ & $2.106\pm0.024$ &   Nan\c{c}ay \\
      UGC 11301 & 18:37:54.5  ~ +17:32:01 &  47.2231 ~   +10.8035 & 4498 & 90.00 & $10.28\pm0.02$ & $ 9.73\pm0.02$ & UH88 & $2.696\pm0.010$ & $2.353\pm0.011$ &     LEDA \\
      UGC 11307 & 18:37:55.3  ~ +27:47:36 &  56.8255 ~   +15.0121 & 3145 & 70.75 & $12.59\pm0.02$ & $12.30\pm0.03$ & UH88 & $2.393\pm0.026$ & $2.011\pm0.029$ &   Nan\c{c}ay \\
       NGC 6674 & 18:38:33.9  ~ +25:22:30 &  54.5925 ~   +13.9190 & 3428 & 56.24 & $ 9.36\pm0.02$ & $ 9.17\pm0.02$ & 2MAS & $2.728\pm0.017$ & $2.386\pm0.019$ &     Mean \\
      UGC 11314 & 18:40:10.4  ~ +21:29:39 &  51.1159 ~   +11.9908 & 4227 & 69.30 & $12.80\pm0.03$ & $12.48\pm0.03$ & UH88 & $2.435\pm0.028$ & $2.056\pm0.026$ &     LEDA \\
      UGC 11320 & 18:40:48.1  ~ +23:41:02 &  53.2171 ~   +12.7669 & 4810 & 90.00 & $10.50\pm0.03$ & $10.05\pm0.03$ & UH88 & $2.696\pm0.005$ & $2.350\pm0.006$ & Springob \\
      UGC 11323 & 18:40:57.9  ~ +23:05:18 &  52.6770 ~   +12.4885 & 4116 & 75.23 & $12.25\pm0.03$ & $11.90\pm0.04$ & UH88 & $2.447\pm0.049$ & $2.072\pm0.055$ &   Nan\c{c}ay \\
    CGMW5-06881 & 18:42:59.7  ~ +21:36:18 &  51.4990 ~   +11.4415 & 4380 & 76.81 & $13.23\pm0.04$ & $12.83\pm0.06$ & IRSF & $2.377\pm0.025$ & $1.993\pm0.028$ &   Nan\c{c}ay \\
      UGC 11344 & 18:44:15.4  ~ +24:08:32 &  53.9790 ~   +12.2362 & 3831 & 62.11 & $10.46\pm0.03$ & $10.24\pm0.03$ & UH88 & $2.656\pm0.015$ & $2.307\pm0.015$ &     LEDA \\
      UGC 11346 & 18:44:16.8  ~ +25:15:23 &  55.0207 ~   +12.6905 & 4645 & 78.46 & $12.63\pm0.02$ & $12.18\pm0.07$ & UH88 & $2.345\pm0.048$ & $1.958\pm0.054$ &   Nan\c{c}ay \\
IRAS 18421+1218 & 18:44:31.2  ~ +12:21:52 &  43.2150 ~   + 7.1160 & 4649 & 72.21 & $10.84\pm0.03$ & $10.35\pm0.04$ & 2MAS & $2.662\pm0.038$ & $2.312\pm0.043$ &   Nan\c{c}ay \\
IRAS 18428+2144 & 18:44:57.7  ~ +21:47:26 &  51.8669 ~   +11.1054 & 4389 & 64.99 & $12.73\pm0.08$ & $12.43\pm0.08$ & 2MAS & $2.507\pm0.030$ & $2.139\pm0.034$ &   Nan\c{c}ay \\
       NGC 6700 & 18:46:04.4  ~ +32:16:46 &  61.8159 ~   +15.1369 & 4582 & 48.45 & $10.21\pm0.04$ & $10.05\pm0.04$ & 2MAS & $2.712\pm0.017$ & $2.383\pm0.018$ &     LEDA \\
      UGC 11355 & 18:47:57.0  ~ +22:56:33 &  53.2278 ~   +10.9655 & 4360 & 66.42 & $10.51\pm0.03$ & $10.20\pm0.03$ & 2MAS & $2.685\pm0.036$ & $2.338\pm0.040$ &   Nan\c{c}ay \\
      UGC 11370 & 18:51:43.5  ~ +26:33:17 &  56.9387 ~   +11.7035 & 4525 & 66.42 & $11.95\pm0.03$ & $11.64\pm0.03$ & 2MAS & $2.542\pm0.020$ & $2.178\pm0.022$ &     Mean \\
IRAS 18575+1845 & 18:59:47.2  ~ +18:49:49 &  50.6935 ~   + 6.6900 & 4689 & 82.22 & $10.50\pm0.02$ & $ 9.76\pm0.05$ & UH88 & $2.749\pm0.018$ & $2.410\pm0.020$ &     Mean \\
      UGC 11394 & 19:03:36.2  ~ +27:36:21 &  59.0506 ~   + 9.7554 & 4232 & 90.00 & $11.06\pm0.03$ & $10.54\pm0.03$ & UH88 & $2.575\pm0.009$ & $2.237\pm0.010$ &     LEDA \\
      UGC 11426 & 19:18:21.7  ~ +34:50:13 &  67.0505 ~   +10.0705 & 4430 & 51.65 & $10.81\pm0.02$ & $10.61\pm0.02$ & 2MAS & $2.563\pm0.028$ & $2.202\pm0.031$ &     Mean \\
      UGC 11565 & 20:28:02.9  ~ +04:57:43 &  49.0593 ~ $-$18.8657 & 3099 & 90.00 & $12.08\pm0.08$ & $11.65\pm0.08$ & 2MAS & $2.360\pm0.050$ & $1.995\pm0.056$ &     LEDA \\
      UGC 11568 & 20:28:19.0  ~ +10:45:24 &  54.2615 ~ $-$15.8773 & 4226 & 75.23 & $10.07\pm0.03$ & $ 9.72\pm0.04$ & 2MAS & $2.689\pm0.015$ & $2.331\pm0.012$ &     LEDA \\
 MCG -02-52-006 & 20:29:35.1 ~$-$12:30:12 &  32.4273 ~ $-$27.3580 & 3932 & 60.67 & $10.53\pm0.03$ & $10.36\pm0.03$ & 2MAS & $2.599\pm0.034$ & $2.242\pm0.038$ &   Nan\c{c}ay \\
 MCG -01-52-016 & 20:38:54.7 ~$-$05:38:24 &  40.4796 ~ $-$26.4290 & 3902 & 75.23 & $10.01\pm0.03$ & $ 9.71\pm0.04$ & 2MAS & $2.595\pm0.017$ & $2.237\pm0.019$ &     LEDA \\
      UGC 11961 & 22:14:46.4  ~ +42:10:53 &  94.3648 ~ $-$11.8375 & 4201 & 80.24 & $10.70\pm0.03$ & $10.20\pm0.08$ & 2MAS & $2.639\pm0.016$ & $2.297\pm0.018$ &     LEDA \\
       NGC 7264 & 22:22:13.8  ~ +36:23:13 &  92.1805 ~ $-$17.4223 & 4273 & 90.00 & $10.42\pm0.02$ & $ 9.98\pm0.02$ & 2MAS & $2.719\pm0.026$ & $2.394\pm0.029$ &     LEDA \\
\hline
\end{tabular}
\label{tab:nir}
\end{table*}
\end{landscape}

}

\begin{table*}[ht]
\caption{Distances and peculiar velocities for the `unbiased' Local void
 sample galaxies.}
\begin{tabular}{rcrrrrcrrrr}
\hline \hline
 & & \multicolumn{4}{c}{UH88/IRSF/2MASS\tablefootmark{a}} 
 & & \multicolumn{4}{c}{2MASS\tablefootmark{b}} \\
\cline{3-6}  \cline{8-11}
\multicolumn{1}{c}{Name} & 
$V_{\mathrm corr}$\tablefootmark{c} & 
\multicolumn{1}{c}{$D$\tablefootmark{d}} & 
\multicolumn{1}{c}{$V_{p}$} & 
\multicolumn{1}{c}{$D^\ast$\tablefootmark{e}} & 
\multicolumn{1}{c}{$V_{p}^\ast$\tablefootmark{f}} && 
\multicolumn{1}{c}{$D$\tablefootmark{d}} & 
\multicolumn{1}{c}{$V_{p}$} & 
\multicolumn{1}{c}{$D^\ast$\tablefootmark{e}} & 
\multicolumn{1}{c}{$V_{p}^\ast$\tablefootmark{f}} \\
\hline
      UGC 11246 & 4287 & $83.19^{+14.15}_{-12.09}$ & $-1535^{+  846}_{ -990}$ & \multicolumn{1}{c}{--} & \multicolumn{1}{c}{--} && $82.01^{+13.91}_{-11.89}$ & $-1452^{+  916}_{ -889}$ & \multicolumn{1}{c}{--} & \multicolumn{1}{c}{--} \\
      UGC 11301 & 4715 & $62.65^{+10.24}_{-8.80}$ & $  331^{+  616}_{ -717}$ & \multicolumn{1}{c}{--} & \multicolumn{1}{c}{--} && $62.10^{+10.15}_{-8.72}$ & $  369^{+  649}_{ -673}$ & \multicolumn{1}{c}{--} & \multicolumn{1}{c}{--} \\
       NGC 6674 & 3656 & $55.00^{+ 9.75}_{-8.28}$ & $ -194^{+  580}_{ -682}$ & \multicolumn{1}{c}{--} & \multicolumn{1}{c}{--} && $55.00^{+ 9.75}_{-8.28}$ & $ -194^{+  580}_{ -682}$ & \multicolumn{1}{c}{--} & \multicolumn{1}{c}{--} \\
      UGC 11320 & 5018 & $73.14^{+11.39}_{-9.83}$ & $ -101^{+  701}_{ -812}$ & $71.80 $ & $   -8^{+  688}_{ -797}$ && $72.60^{+11.29}_{-9.75}$ & $  -63^{+  695}_{ -805}$ & $71.28$ & $   30^{+  720}_{ -753}$ \\
      UGC 11344 & 4040 & $65.97^{+11.21}_{-9.58}$ & $ -578^{+  671}_{ -785}$ & \multicolumn{1}{c}{--} & \multicolumn{1}{c}{--} && $61.03^{+10.36}_{-8.86}$ & $ -227^{+  971}_{ -375}$ & \multicolumn{1}{c}{--} & \multicolumn{1}{c}{--} \\
IRAS 18421+1218 & 4850 & $75.56^{+17.72}_{-14.17}$ & $ -439^{+ 1061}_{-1327}$ & $70.62 $ & $  -93^{+  992}_{-1240}$ && $75.56^{+17.72}_{-14.17}$ & $ -439^{+ 1061}_{-1327}$ & $70.62$ & $  -93^{+  992}_{-1240}$ \\
       NGC 6700 & 4774 & $81.73^{+14.39}_{-12.23}$ & $ -946^{+  856}_{-1007}$ & \multicolumn{1}{c}{--} & \multicolumn{1}{c}{--} && $81.73^{+14.39}_{-12.23}$ & $ -946^{+  857}_{-1007}$ & \multicolumn{1}{c}{--} & \multicolumn{1}{c}{--} \\
      UGC 11355 & 4555 & $73.17^{+17.53}_{-14.14}$ & $ -566^{+  990}_{-1227}$ & \multicolumn{1}{c}{--} & \multicolumn{1}{c}{--} && $73.17^{+17.53}_{-14.14}$ & $ -566^{+  991}_{-1227}$ & \multicolumn{1}{c}{--} & \multicolumn{1}{c}{--} \\
IRAS 18575+1845 & 4859 & $79.60^{+14.44}_{-12.22}$ & $ -713^{+  855}_{-1011}$ & \multicolumn{1}{c}{--} & \multicolumn{1}{c}{--} && $70.26^{+12.82}_{-10.84}$ & $  -50^{+ 1422}_{ -234}$ & \multicolumn{1}{c}{--} & \multicolumn{1}{c}{--} \\
      UGC 11426 & 4620 & $65.12^{+10.80}_{-8.93}$ & $   63^{+  788}_{ -953}$ & $51.66 $ & $ 1005^{+  625}_{ -756}$ && $65.12^{+10.80}_{-8.93}$ & $   63^{+  788}_{ -953}$ & $51.66$ & $ 1005^{+  625}_{ -756}$ \\
      UGC 11568 & 4288 & $57.01^{+ 9.46}_{-8.11}$ & $  298^{+  568}_{ -662}$ & \multicolumn{1}{c}{--} & \multicolumn{1}{c}{--} && $57.01^{+ 9.46}_{-8.11}$ & $  298^{+  567}_{ -662}$ & \multicolumn{1}{c}{--} & \multicolumn{1}{c}{--} \\
 MCG -02-52-006 & 3990 & $56.23^{+12.54}_{-10.17}$ & $   55^{+  742}_{ -914}$ & $53.97 $ & $  213^{+  712}_{ -877}$ && $56.23^{+12.54}_{-10.17}$ & $   55^{+  742}_{ -914}$ & $53.97$ & $  213^{+  712}_{ -878}$ \\
 MCG -01-52-016 & 3953 & $42.05^{+ 6.99}_{-5.93}$ & $ 1009^{+  446}_{ -525}$ & $39.17 $ & $ 1211^{+  415}_{ -489}$ && $42.05^{+ 6.99}_{-5.93}$ & $ 1009^{+  446}_{ -525}$ & $39.17$ & $ 1211^{+  415}_{ -490}$ \\
      UGC 11961 & 4223 & $64.00^{+11.20}_{-9.49}$ & $ -256^{+  684}_{ -807}$ & $62.20 $ & $ -130^{+  665}_{ -784}$ && $64.00^{+11.20}_{-9.49}$ & $ -256^{+  684}_{ -807}$ & $62.20$ & $ -130^{+  665}_{ -784}$ \\
       NGC 7264 & 4231 & $82.58^{+16.77}_{-13.94}$ & $-1549^{+  976}_{-1174}$ & \multicolumn{1}{c}{--} & \multicolumn{1}{c}{--} && $82.58^{+16.77}_{-13.94}$ & $-1549^{+  976}_{-1173}$ & \multicolumn{1}{c}{--} & \multicolumn{1}{c}{--} \\
\hline
\end{tabular}
\tablefoot{ 
\tablefoottext{a}{These values are based on $H$-band photometry from UH88 and
 IRSF observations when they are available and on 2MASS XSC otherwise.}
\tablefoottext{b}{These values are based on $H$-band photometry from
 2MASS XSC alone.}
\tablefoottext{c}{See Sect. 4.1.} 
\tablefoottext{d}{Distances from IRTFR with bias corrections (when its
 normalized distance moduli $X>-0.5$; see Sect. 4.2.2 for details).}
\tablefoottext{e}{Distances before the correction for bias for galaxies
 with $X>-0.5$.} 
\tablefoottext{f}{Peculiar velocity without correction for bias (i.e.,
 using $D^\ast$ instead of $D$).} 
}
\label{tab:vpec}
\end{table*}


\clearpage

\end{document}